\documentclass[journal=AELCCP,manuscript=letter]{achemso}

\usepackage[version=3]{mhchem} % Formula subscripts using \ce{}
\usepackage{paracol}
\usepackage{graphicx}
\usepackage[dvipsnames]{xcolor}
\author{Jakub Pawelko}
\affiliation{Univ Rennes, CNRS, Institut des Sciences Chimiques de Rennes - UMR 6226, F-35000 Rennes, France}
\author{Xavier Rocquefelte}
\affiliation{Univ Rennes, CNRS, Institut des Sciences Chimiques de Rennes - UMR 6226, F-35000 Rennes, France}
\email{xavier.rocquefelte@univ-rennes.fr}
\author{Auguste Tetenoire}
\affiliation{Laboratory of Computational Chemistry and Biochemistry, École Polytechnique Fédérale de Lausanne, Lausanne, Switzerland}
\author{David Le Coq}
\affiliation{Univ Rennes, CNRS, Institut des Sciences Chimiques de Rennes - UMR 6226, F-35000 Rennes, France}
\author{Laurent Calvez}
\affiliation{Univ Rennes, CNRS, Institut des Sciences Chimiques de Rennes - UMR 6226, F-35000 Rennes, France}
\author{Eric Furet}
\affiliation{Univ Rennes, CNRS, Institut des Sciences Chimiques de Rennes - UMR 6226, F-35000 Rennes, France}
\email{eric.furet@enscr.fr}

\title{Mechanistic Insights into Li\textsuperscript{+} Transport Enabled by Isolated Sulfur Species in Li\textsubscript{3}PS\textsubscript{4} Glasses}

\abbreviations{}

\begin{document}

\begin{abstract}
All-solid-state lithium-ion batteries have renewed interest in high-performance solid electrolytes. Li\textsubscript{3}PS\textsubscript{4} (Li\textsubscript{2}S–P\textsubscript{2}S\textsubscript{5}) glasses are among the most studied due to their high ionic conductivity, traditionally ascribed to rotational motion of polyhedral units facilitating Li\textsuperscript{+} migration. Using \textit{ab initio} molecular dynamics, we investigate Li-ion diffusion in Li\textsubscript{3}PS\textsubscript{4} glass, demonstrating that our structural model reproduces experimental neutron and X-ray diffraction patterns and conductivity measurements. Importantly, we identify a previously unrecognized diffusion mechanism: Li\textsuperscript{+} ions near isolated sulfur species (S\textsubscript{n} with n = 1, 3) display significantly enhanced mobility, with atomic displacements up to 1.7 greater than those associated with bulkier polyhedral units. These results highlight the critical role of free sulfur species in promoting fast ionic transport, providing insights for the rational design of glass compositions with optimized conductivity for solid-state battery applications.
\end{abstract}


%% Start d'intro %%

\hspace{0.5cm}Lithium batteries play a crucial role in modern daily life, powering the electronic devices used worldwide. Most current lithium-ion batteries employ liquid electrolytes\cite{Armand2008,Famprikis2019}. Incorporating solid-state electrolytes (SSE)\cite{Choi2024, Joshi2025, Sung2024} into these systems offers the potential to reduce device size and enhance safety by hindering dendrite growth between the electrodes.\cite{Quartarone2011} However, solid-state electrolytes remain limited by characteristics inherent to their nature, most notably their lower ionic conductivity in comparison to liquid electrolytes.\cite{Bruce2012} The Li–P–S system represents one of the most promising compositions for solid-state electrolytes among other ionic superconductors, as both its crystalline and amorphous phases exhibit relatively high ionic conductivity.\cite{Wang2015,Jun2024_2} When it comes to these systems, glassy electrolytes offer several advantages over ceramic counterparts, including isotropic transport properties, greater chemical flexibility and plasticity, and improved compatibility with electrode materials\cite{Huang2025}. Additionally, Li\textsubscript{3}PS\textsubscript{4} glass is one of the most widely studied amorphous materials whose structural properties are well documented\cite{Dietrich2017,Lee2023,Ohara2016, Smith2020}. 

Although substantial progress has been made in rationalizing the structural and topological features governing fast Li-ion transport in crystalline superionic conductors \cite{Jun2024_2}, a comparable understanding in glasses remains elusive due to their structural complexity and lack of periodicity. Li\textsubscript{3}PS\textsubscript{4} demonstrates this challenge: although its amorphous phase (3Li\textsubscript{2}S:1P\textsubscript{2}S\textsubscript{5}) exhibits high Li-ion conductivity ($10^{-4}$--$10^{-3}$~S\,cm$^{-1}$)\cite{Dietrich2017,Hayashi2001}, crystalline $\gamma$-Li\textsubscript{3}PS\textsubscript{4} is orders of magnitude less conductive ($10^{-8}$--$10^{-7}$~S\,cm$^{-1}$)\cite{Tachez1984}. This contrast has been ascribed to the coexistence of multiple polyanionic species in the glass, namely PS\textsubscript{4}\textsuperscript{3-}, P\textsubscript{2}S\textsubscript{6}\textsuperscript{4-}, and P\textsubscript{2}S\textsubscript{7}\textsuperscript{4-}\cite{Dietrich2017,Ohara2016}, as well as to the possible influence of their rotational dynamics (the so-called paddle-wheel effect)\cite{Smith2020}.

However, the paddle-wheel mechanism \cite{Lunden1988} remains controversial \cite{Jansen1991,Jun2024}. Alternative interpretations, including the billiard-ball mechanism based on collision-mediated cascade processes\cite{Fang2022,Baba2024}, have been proposed. Recent theoretical work employing quaternion-based rotational event detection found no spatial or temporal correlation between anion-group rotations and Li-ion hopping, thereby excluding the paddle-wheel mechanism and instead attributing fast transport to correlated static disorder in Li-site occupancy and anion-group orientations, termed the soft-cradle effect\cite{Jun2024_2}. This concept is closely related to the facilitation of cation diffusion by wiggling motion and translational flexibility of weakly bound anion groups \cite{Gupta2021}. Importantly, the possible role of other units, such as free sulfur species, was not addressed in this framework, despite their observation in molecular dynamics simulations\cite{Lee2023}. In parallel, studies of sodium and sulfide glasses suggest that isolated sulfur species can promote preferential conduction pathways \cite{Kassem2020,Kassem2022}.

In this paper, we computationally investigate Li transport in Li\textsubscript{3}PS\textsubscript{4} glass with a focus on the transport mechanisms associated with its diverse structural species. Particular attention is devoted to homonuclear sulfur species (S\textsubscript{n}, (n = 1–3)), whose role in lithium-based glasses has not yet been explored.
\begin{figure}[!b]
    \centering
    \includegraphics[scale=0.27]{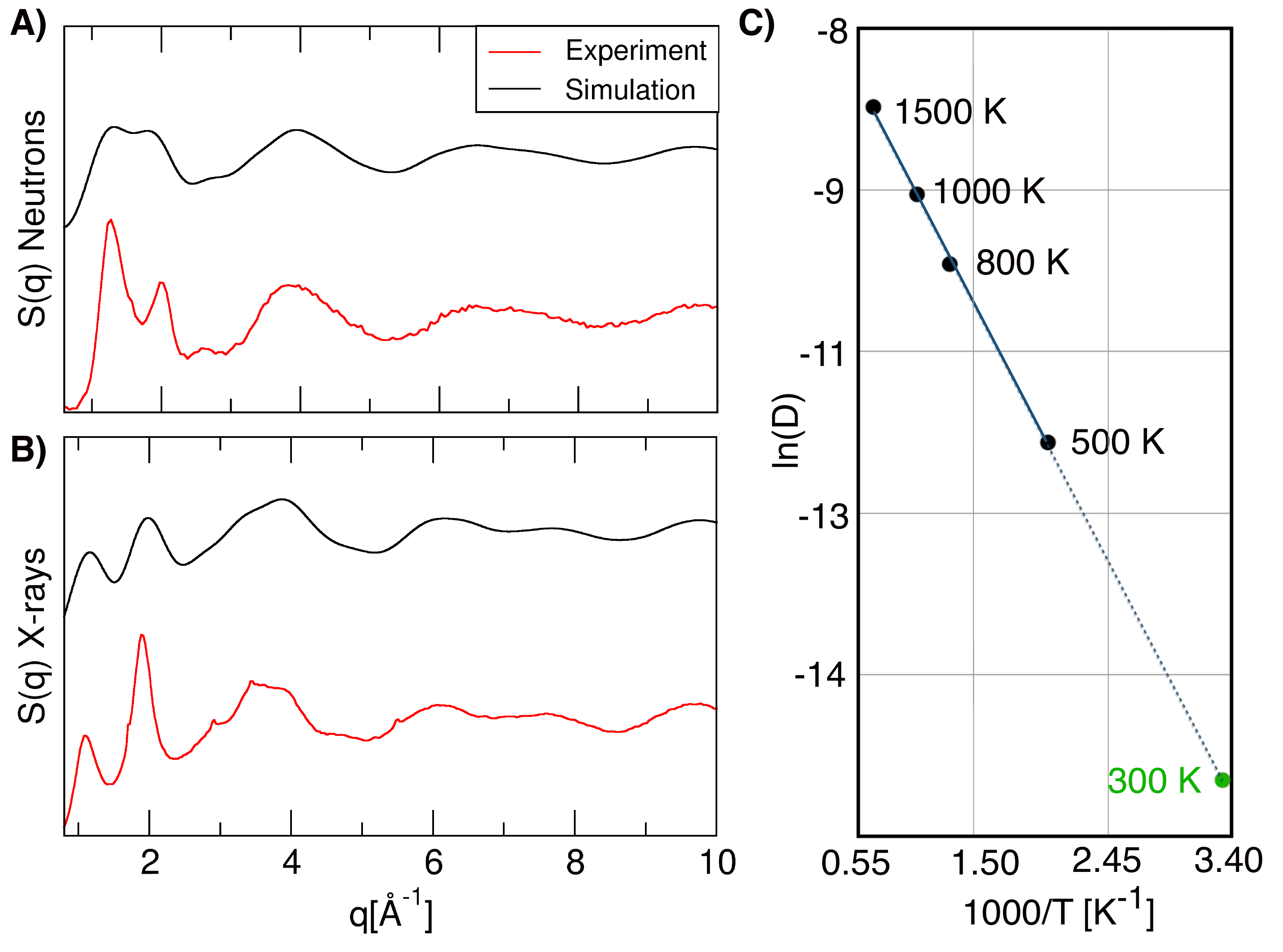}
    \caption{(\textbf{A, B}) Neutron (\textbf{A}) and X-ray (\textbf{B}) structure factors of the simulated Li\textsubscript{3}PS\textsubscript{4} glass model (black) at 300 K compared with experimental data (red). (\textbf{C}) Arrhenius plot averaged over four independent molecular dynamics simulations at liquid-state temperatures, extrapolated to 300 K (green). The linear regression, $\ln(D) = 2640.2/T - 6.6078$ ($R^2 = 0.9988$), was used to determine the lithium ions diffusion coefficient and conductivity.}    
    \label{fig:figure1}
\end{figure}

As shown in \textbf{Figures 1A and 1B}, the simulated neutron and X-ray structure factors closely match the experiment\cite{Ohara2016}, including subtle features such as the third neutron peak at 2.5~\AA$^{-1}$. Minor sharp peaks at 3.0, 3.2, and 5.6~\AA$^{-1}$ in the experimental X-ray $S(Q)$ suggest slight crystallization of the sample. This agreement, together with the accurate simulation of the Raman spectrum previously reported\cite{Pawelko2025}, using our recently developed theoretical approach, confirms that our \textit{in silico} model reproduce both the global and local glass structures. Building on this validated model, lithium diffusion was calculated at 1500, 1000, 800, and 500 K (\textbf{Figure 1C}) equivalent to liquid state of the glass. Linear Arrhenius extrapolation, yields a diffusion coefficient of $D \approx 2.0 \times 10^{-7}$~cm$^{2}$\,s$^{-1}$ at 300~K, which is in good agreement with previous calculations \cite{Smith2020}. The corresponding ionic conductivity calculated by means of Nernst-Einstein relation\cite{Einstein1905,Verma2024} (see Supp. Mat.) is $19.7 \times 10^{-3}$~S\,cm$^{-1}$, close to the value of $19.0 \times 10^{-3}$~S\,cm$^{-1}$ reported in prior simulations\cite{Smith2020}. In contrast, experimental values are approximately two orders of magnitude lower ($\sim 2.8$--$3.0 \times 10^{-4}$~S\,cm$^{-1}$) \cite{Dietrich2017,Hayashi2001}, reflecting that our calculations do not account for defects and grain boundaries in the electrolyte.
\begin{figure}[!t]
    \centering
    \includegraphics[scale=0.4]{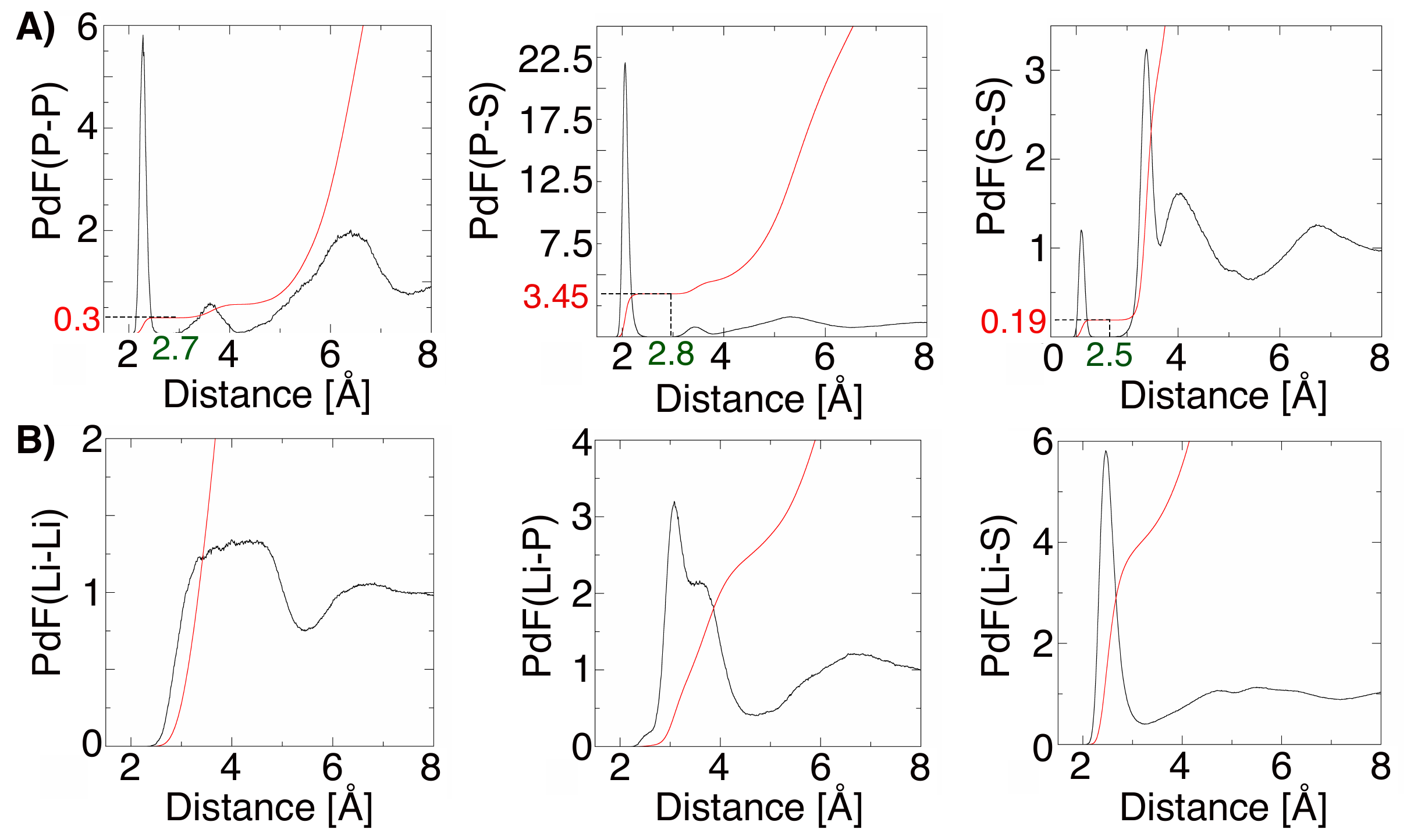}
    \caption{Partial pair distribution functions (black) and their cumulative integrals (red) averaged over four independent molecular dynamics simulations. (\textbf{A}) P–P, P–S, and S–S correlations; the plateau of each integration is marked with a dashed line indicating the average coordination, and the cut-off radius r\textsubscript{cut} is shown in green. (\textbf{B}) Li–Li, Li–P, and Li–S correlations.}    
    \label{fig:figure2}
\end{figure}

\textbf{Figure 2A} displays the P-S, P-P and S-S partial pair distribution functions (PDFs). In each case, the sharp peak allows an unambiguous determination of the average first coordination number. \textbf{Table 1} summarizes these values along with the corresponding cutoff radii (r\textsubscript{cut}), which are also indicated on the PDFs in \textbf{Figure 2}.
\begin{table}[!b]
\centering
\caption{Average coordination numbers, $n(r_\mathrm{cut})$, and corresponding cut-off radii, $r_\mathrm{cut}$, for P and S environments in Li\textsubscript{3}PS\textsubscript{4} glass.}
\label{tab:my-table1}
\begin{tabular}{|l|c|c|c|}
\hline
                                      & \textbf{P-P}  & \textbf{P-S}  & \textbf{S-S}  \\ \hline
$\pmb{n(r_\mathrm{cut})}$    & 0.30 & 3.45 & 0.19 \\ 
$\pmb{r_\mathrm{cut}}$ \textbf{(Å)} & 2.7  & 2.8  & 2.5  \\ \hline
\end{tabular}
\end{table}
On average, each P atom is bonded to 3.75 neighbors (3.45 S and 0.30 P) atoms, close to the expected quadricoordination mode. This indicates that the glass network is predominantly composed of fourfold-coordinated phosphorus species, with a minor fraction of homopolar P-P and S-S bonds, characterized by n(r\textsubscript{cut}) = 0.30 and 0.19, respectively.

Analysis of lithium coordination is presented in \textbf{Figure 2B}, showing Li-X (X = Li, S, P) PDFs. No Li-Li distances are consistent with purely covalent or iono-covalent bonds (r\textsubscript{cov}(Li) = 1.33 Å) and the first peak is shallow, spanning from 2.5 to 5.0 Å. In contrast, the Li-S PDF exhibits a well-defined first peak, corresponding to an average of nearly 4 neighbors, while the Li-P PDF shows an intermediate pattern with a small shoulder at $\sim$2.6 Å and two merged peaks. The latter peaks correspond to Li in the vicinity of S within P\textsubscript{m}X\textsubscript{n} polyhedra, while the shoulder is likely due to undercoordinated P atoms. 

\begin{figure}[!t]
    \centering
    \includegraphics[scale=0.3]{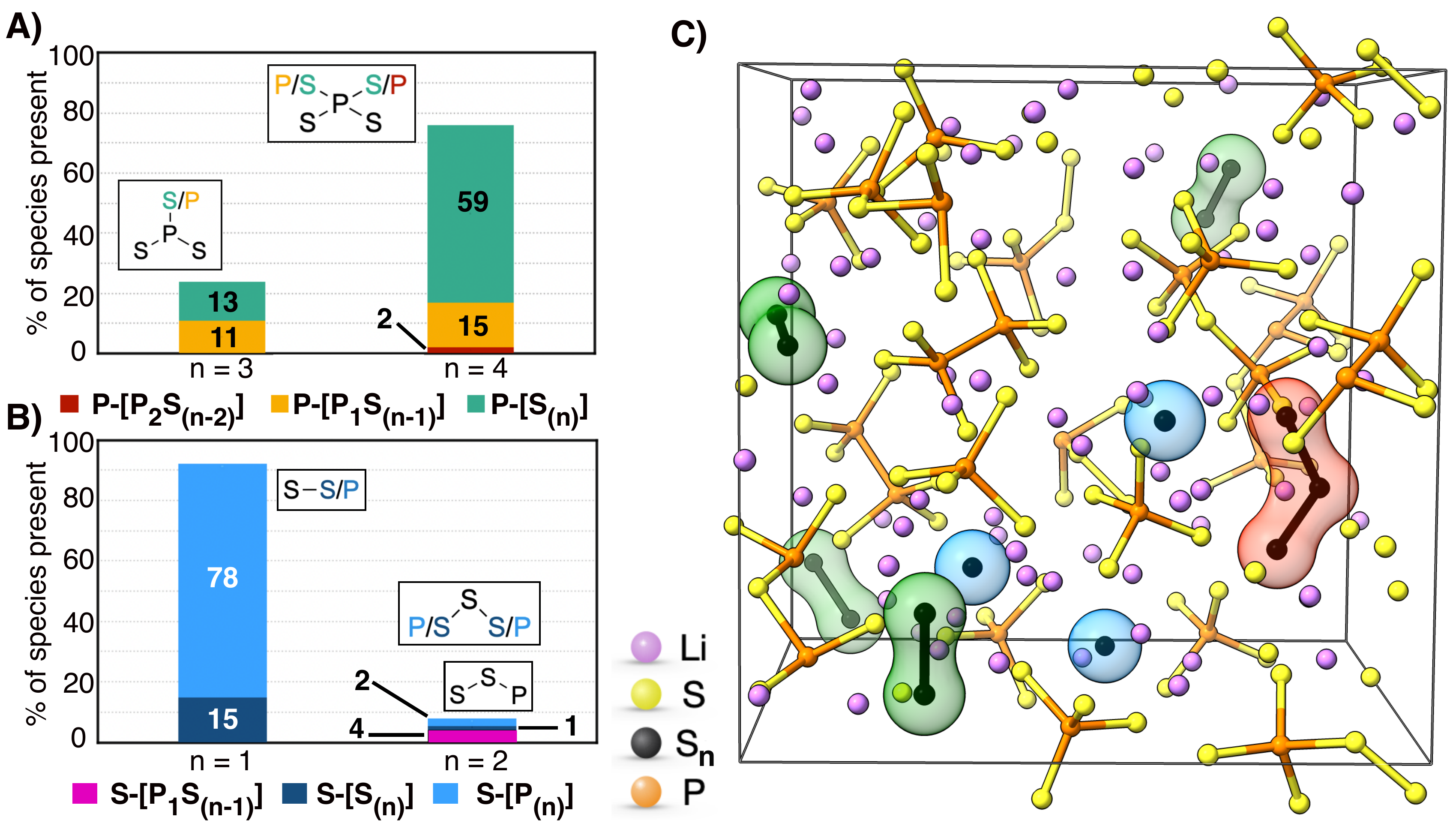}
        \caption{Connectivity distributions and snapshots of Li\textsubscript{3}PS\textsubscript{4} glass. (\textbf{A, B}) Phosphorus-(\textbf{A}) and sulfur-(\textbf{B}) resolved coordination environments. Key structural connections are color-coded: P–S (green), P–P–S (yellow), P–(P\textsubscript{2})–S (red), S–P (light blue), S–S (dark blue), and S–S–P (black). (\textbf{C}) Snapshot of the glass at 300 K from MD-1, highlighting three types of sulfur ions (S\textsubscript{1}, blue; S\textsubscript{2}, green and S\textsubscript{3}, red).
        }
    \label{fig:figure3}
\end{figure}
The cut-off radii from \textbf{Table 1} were used to perform a statistical analysis of S and P coordination environments. The values are shown as histograms in \textbf{Figure 3A} (P environment) and \textbf{Figure 3B} (S environment), providing a refined view of the structural units in the glass. For phosphorus, $\sim$76\% of atoms are quadricoordinated and $\sim$24\% tricoordinated, with similar fractions of P-P bonds (11\% and 17\%). These results indicate a predominance of PS\textsubscript{4} tetrahedra, along with some P\textsubscript{2}S\textsubscript{6} (S\textsubscript{3}P-PS\textsubscript{3}) units. The sulfur histograms reveal that $\sim$2\% of S atoms are P-S-P species, corresponding to P\textsubscript{2}S\textsubscript{7} (S\textsubscript{3}P-S-PS\textsubscript{3}) units. 

Moreover, smaller S\textsubscript{n} species (n = 1–3) that are not fully captured in the coordination histograms are visible in snapshots of the glass at 300 K, as illustrated in \textbf{Figure 3C}. These species are not covalently bonded to atoms other than sulfur and form only ionic interactions with lithium. Averaged over four independent trajectories, the simulation cells contain 2.5 S\textsubscript{1}, 3.0 S\textsubscript{2}, and 1.5 S\textsubscript{3} sulfur species (\textbf{Figure S1)}. To establish their anionic nature, Mulliken charges\cite{Mulliken1955,Csizmadia1976} were computed for all atoms in a 300 K snapshot. While absolute magnitudes of atomic charges have limited physical meaning, relative values provide valuable insights. Mulliken charge analysis reveals that lithium, phosphorus, and sulfur atoms (excluding S\textsubscript{n} units) exhibit charges ranging from +0.4 to +0.6, +0.2 to +0.6, and -0.6 to -0.2, respectively. Few notable exceptions, can be observed:  S\textsubscript{1} carries a charge of -0.8, while the central atom of S\textsubscript{3} shows a significantly increased charge of -0.1. These values confirm the expected ionic character of the glass: Li and P act as cations, while S serves as an anion. An alternative way to estimate the charge state of S\textsubscript{2} and S\textsubscript{3} clusters is by comparing their geometrical parameters with those of the corresponding entities in the gas phase. Our analysis shows that the geometrical parameters of both S\textsubscript{2} and S\textsubscript{3} clusters in the glass, across the 4 MDs, are consistent with charge state in between -2 and -1 (Figure S2), hence confirming that all S\textsubscript{n} species are anionic. 

To further unravel diffusion mechanisms in the superionic glass, we characterize Li-ion migration events using the functional h\textsubscript{i}\cite{Smith2020,Fang2022}, which identifies long-lived Li displacements exceeding a distance $a$ over a time interval $t$ (see Supp. Mat. for methodological details). A minimal displacement of $a = 1.6$ Å was adopted, consistent with previous studies \cite{Burbano2016,Smith2020}. \textbf{Figure 4A} shows the Li-ion diffusion events identified at 500 K over a 12 ps time window. This temperature provides sufficient atomic mobility to probe the Li-ions dynamics while largely preserving structural integrity. Among the seven identified fast Li-ions, four are located in close proximity to isolated sulfur species. Specifically, Li(16) and Li(25) are near S\textsubscript{1}, while Li(66) and Li(67) are near S\textsubscript{3}. Unlike most Li-S environments, Li(16) is tricoordinated by sulfur and replaces another Li-ion with a similar coordination (Li(27)), which subsequently occupies the position of Li(55) (\textbf{Figure 4B}). The atypical LiS\textsubscript{3} coordination generates preferential migration pathways within the glass network, analogous to those proposed for other glass families \cite{Kassem2020,Kassem2022}.However, here instead of enlarged Li sites, \cite{Jun2024_2} we evidenced the impact of diffusion channels formed between under-coordinated LiS\textsubscript{3} units adopting planar triangular geometries. The cooperative motion of Li-ions in the vicinity of monosulfur species (S\textsubscript{1}) is further quantified by mean-squared displacement (MSD) analysis (\textbf{Figure 4C}), which shows similar displacement profiles for successive Li\textsuperscript{+} ions, suggesting a cooperative transport mechanism.  

\begin{figure}
    \centering
    \includegraphics[scale=0.4]{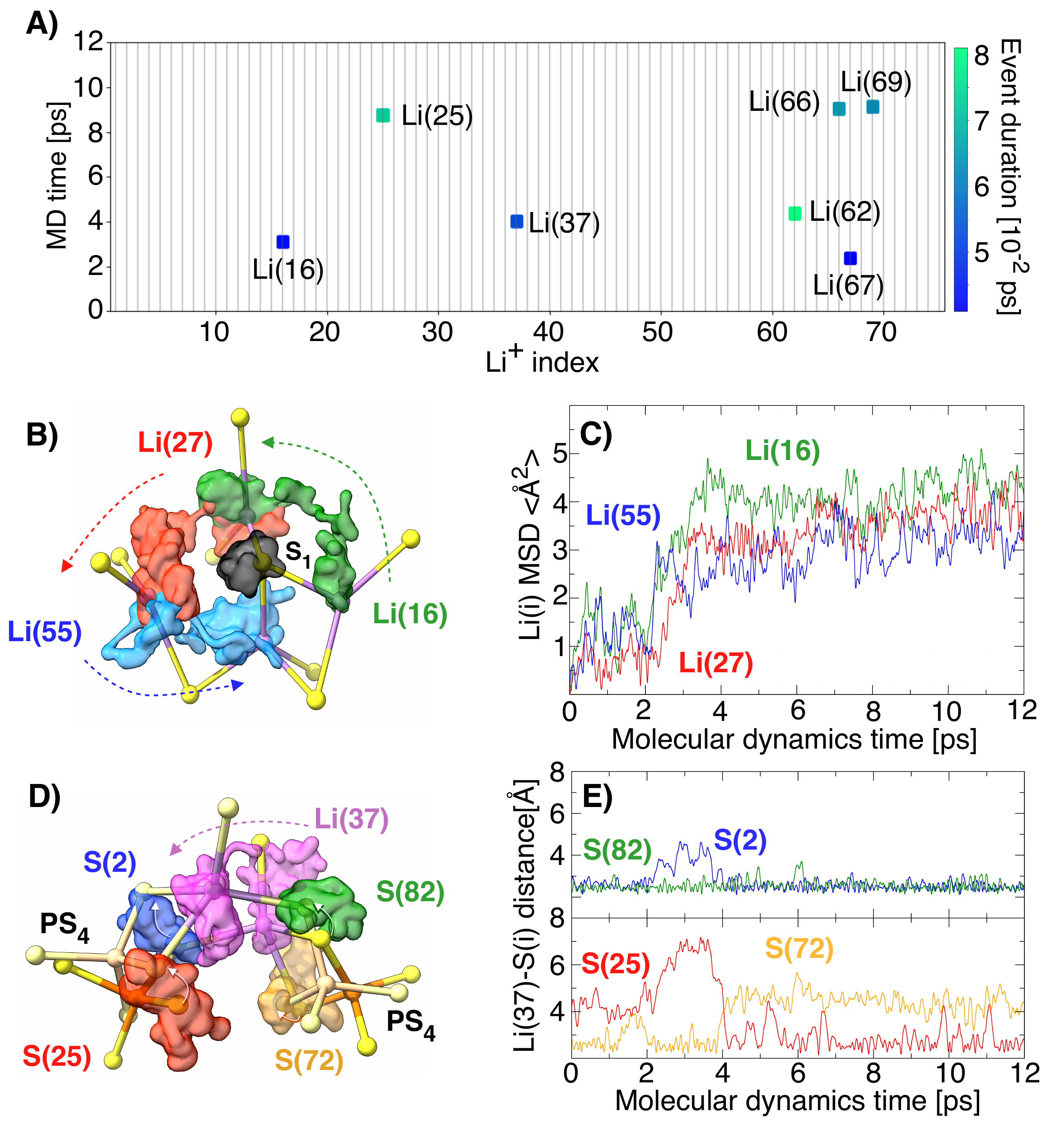}
        \caption{Li\textsuperscript{+} migration events over 12 ps at 500 K. (\textbf{A}) Identification of ions participating in migration events and event durations for MD-1. (\textbf{B}) Trajectories of cooperative Li ions (Li(16), green; Li(27), red; Li(55), blue) near a monosulfur species (S\textsubscript{1}, black); arrows indicate displacement directions. (\textbf{C}) Mean-squared displacement of Li ions involved in cooperative motion shown in (B).  (\textbf{D}) Trajectory of Li(37) (pink) and the four sulfur atoms of a PS\textsubscript{4} tetrahedron involved in a soft-cradle–like movement (S(2),blue; S(82), green; S(25), red and S(72), orange); arrows indicate movement directions. (\textbf{E}) Distances between Li(37) and surrounding sulfur atoms during soft-cradle–like movement.}
    \label{fig:figure4}
\end{figure}

In contrast, Li-ions located near tetrahedral PS\textsubscript{4} species (e.g., Li(37)) exhibit different transport behavior. In this case, lithium remains tetrahedrally coordinated and undergoes site-to-site migration between neighboring S\textsubscript{4} tetrahedral cages. As shown in \textbf{Figure 4D}, the trajectories of Li(37), S(2) and S(82) are strongly correlated (depicted in pink, blue and green, respectively), resulting in nearly constant Li-S bond lengths over the entire 12 ps trajectory (\textbf{Figure 4E, top}). Meanwhile, the two remaining sulfur atoms defining the tetrahedral environment vary during Li(37) migration (\textbf{Figure 4E, bottom}) by substitution of S(72) with S(25), shown in orange and green, respectively. This behavior is consistent with the soft-cradle effect\cite{Jun2024,Jun2024_2} and is also observed for other lithium ions within the same trajectory (\textbf{Figure S3}) and across all independent molecular dynamics simulations (\textbf{Figure S4}).

To quantitatively assess Li\textsuperscript{+} displacements associated with the mechanisms described above, we calculated total and partial mean squared displacements (t-MSD and p-MSD) at each temperature during the quenching process. Values were averaged over four independent molecular dynamics simulations to ensure better representativeness. The t-MSD corresponds to an average over all Li-ions, whereas the p-MSD is restricted to Li-ions located near specific structural entities in the glass, namely P\textsubscript{x}S\textsubscript{y}, S\textsubscript{1}, S\textsubscript{2} and S\textsubscript{3} species, which are associated with 73, 13, 9 and 5\% of Li-ions in the glass, respectively. \textbf{Figure 5} and \textbf{Table 2} summarize these results. At 300 K, the t-MSD is 262±27 Å\textsuperscript{2} (2$\sigma$), and the p-MSD values are 206±18, 293±84, 172±11 and 345±96 Å\textsuperscript{2} for P\textsubscript{x}S\textsubscript{y}, S\textsubscript{1}, S\textsubscript{2} and S\textsubscript{3} units, respectively. The larger uncertainties obtained for p-MSD associated to S\textsubscript{1} and S\textsubscript{3} species occur primarily from the MD-2 simulation, which deviate significantly from the other three trajectories (\textbf{Table S1}), hence exemplifying the paramount importance of performing several MD to achieve statistically pertinent results.

\begin{figure}[!t]
    \centering
    \includegraphics[scale=0.4]{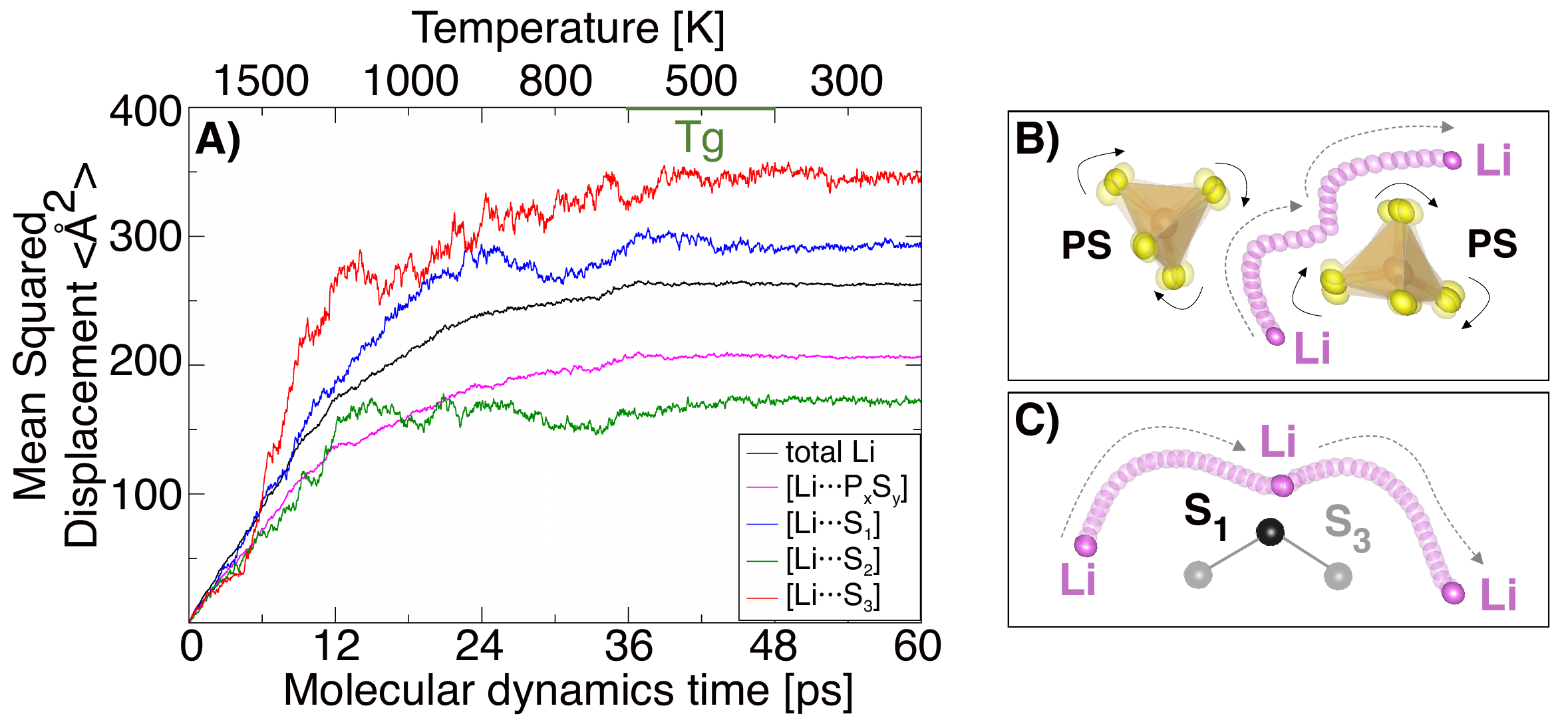}
    \caption{(\textbf{A}) Average mean squared displacement of lithium ions associated with species present in the glass namely S\textsubscript{1} (blue), S\textsubscript{2} (green), S\textsubscript{3} (red) and P associated molecules (pink) compared to total lithium displacement (black). (\textbf{B}) Scheme representing Li-ions movement in proximity to the isolated sulfur species with high diffusion (S\textsubscript{n}, n = 1,3) by the correlated transport of Li ions. (\textbf{C}) Scheme of Li\textsuperscript{+} near the P\textsubscript{x}S\textsubscript{y} species - soft-cradle–like mechanism. Lithium atoms are shown in pink, phosphorus in orange, P bonded sulfur in yellow and S bonded sulfur in black.}
    \label{fig:figure5}
\end{figure}

Notably, all p-MSD values remain below the t-MSD, except for Li-ions located near S\textsubscript{1} and S\textsubscript{3} species, which exhibit p-MSD values approximately 40\% and 70\% larger than those associated with P\textsubscript{x}S\textsubscript{y} units. It should be noted that the lowest p-MSD value is associated to Li-ions near S\textsubscript{2} units, which were also not detected by the h\textsubscript{i} functional as fast ions. Therefore, it appears that only S\textsubscript{1} and S\textsubscript{3} uneven species contribute to improve the ionic conductivity. Furthermore, contrary to earlier reports suggesting that bulky species hinder Li-ion diffusion\cite{Ohara2016}, our results indicate that these species do not significantly impede Li\textsuperscript{+} transport, in agreement with more recent studies\cite{Lee2023}. 

\begin{table}[!b]
\centering
\caption{Partial Mean square displacement (p-MSD) values at 300 K for Li-ions near S\textsubscript{n} (n = 1, 2, 3) and P\textsubscript{x}S\textsubscript{y} species. Reported values correspond to averages over the indicated number of Li ions (N\textsubscript{Li}) and over four independent molecular dynamics simulations. Uncertainties represent standard deviations at the $2\sigma$ confidence level.}
\label{tab:my-table2}
\begin{tabular}{|c|c|c|c|}
\hline
\multicolumn{1}{|l|}{\textbf{Type of species}} & \textbf{N\textsubscript{Li}} & \textbf{\textless{}MSD\textgreater {[}Å\textsuperscript{2}{]}} \\ \hline
\textbf{[Li···S\textsubscript{1}]}                                    & 10                                            & 292.9±83.9                                   \\ \hline
\textbf{[Li···S\textsubscript{2}]}                                    & 7                                             & 172.0±11.3                                   \\ \hline
\textbf{[Li···S\textsubscript{3}]}                                    & 3.25                                          & 345.2±95.8                                   \\ \hline
\textbf{[Li···P\textsubscript{x}S\textsubscript{y}]}                                    & 54.75                                         & 206.2±18.2                                   \\ \hline
\textbf{total Li}                                 & 75                                            & 262.6±26.7                                   \\ \hline
\end{tabular}
\end{table}

In this work, we simulated the atomic structure of Li\textsubscript{3}PS\textsubscript{4} glass using \textit{ab initio} molecular dynamics. The resulting model reproduces experimental neutron and X-ray structure factors as well as ionic conductivity. Comprehensive structural analysis reveals the coexistence of P\textsubscript{x}S\textsubscript{y} (PS\textsubscript{4}, P\textsubscript{2}S\textsubscript{6}, P\textsubscript{2}S\textsubscript{7} and extended P–S chain), and isolated sulfur species (S\textsubscript{n} with n = 1, 2, 3). Lithium transport is shown to be strongly structure-dependent: Li\textsuperscript{+} migration proceeds \textit{via} a soft-cradle–like mechanism near weakly bound PS\textsubscript{4} tetrahedra (\textbf{Figure 5B}), whereas S\textsubscript{1} and S\textsubscript{3} species enable a distinct tunnel-like cooperative transport pathway that supports larger Li displacements (\textbf{Figure 5C}). Quantitatively, Li\textsuperscript{+} ions coordinated to S\textsubscript{1} and S\textsubscript{3} species exhibit mean squared displacements $\sim$1.4 and 1.7 times larger than those associated with P\textsubscript{x}S\textsubscript{y} units. This analysis directly links specific structural motifs to enhanced diffusion and ionic conductivity \textit{via} the Nernst-Einstein relation\cite{Einstein1905,Verma2024}. These results establish free S\textsubscript{1} and S\textsubscript{3} anions as promoting fast-ion transport and suggest controlling their incorporation as an effective design strategy for sulfide glass electrolytes improvement. Specifically, glass compositions and processing routes that stabilize isolated sulfur species without compromising structural integrity are predicted to promote percolating Li-ion pathways and thus higher conductivity. More broadly, this work demonstrates that tailoring the population and connectivity of isolated sulfur species provides a rational strategy for engineering high-performance sulfide solid electrolytes for next-generation solid-state batteries.

\begin{acknowledgement}
\hspace{0.5cm}The authors acknowledge Grand équipement national de calcul intensif (GENCI) for granting access to the High-performance computing (HPC) resources of TGCC (Très grand centre de calcul du CEA), CINES (Centre informatique national de l'enseignement supérieur) and IDRIS (Institut du développement et des ressources en informatique scientifique) networks under the allocation 2025-AD010916827 and 2026-A0190907682.

The authors acknowledge Prof. Koji Ohara (Shimane University, Japan) for providing detailed data of experimental neutrons and X-rays structure factors.
\end{acknowledgement}

\begin{suppinfo}
\hspace{0.5cm} Computational details, including software and simulation protocols; snapshots from four independent molecular dynamics simulations at 300 K highlighting isolated sulfur entities; analysis of connection between S-S bond lengths in S\textsubscript{2} and S\textsubscript{3} entities and their charge as well as detailed identification of Li\textsuperscript{+} ions participating in migration events across all trajectories (MD1 - MD4).
\end{suppinfo} 

%%%%%%%%%%%%%%%%%%%%%%%%%%%%%%%%%%%%%%%%%%%%%%%%%%%%%%%%%%%%%%%%%%%%%
%% The a$ \bibliography command should be placed here.
\bibliography{achemso-demo}
\end{document}